\documentclass[acmtog,nonacm]{acmart}
\AtBeginDocument{%
  }

\usepackage{bm}
\usepackage{color}

\usepackage{amssymb}
\usepackage{enumitem}

\begin{document}
\definecolor{revised}{rgb}{0,0,1}
\DeclareRobustCommand{\rev}[1]{{#1}}
\title{DeMapGS: Simultaneous Mesh Deformation and Surface Attribute Mapping via Gaussian Splatting}


\author{Shuyi Zhou}
\email{zhoushuyi495@cvl.iis.u-tokyo.ac.jp}
\affiliation{%
  \institution{The University of Tokyo}
  \city{Tokyo}
  \country{Japan}
}

\author{Shengze Zhong}
\email{zhong\_shengze@cyberagent.co.jp}
\affiliation{%
  \institution{CyberAgent}
  \city{Tokyo}
  \country{Japan}
}

\author{Kenshi Takayama}
\email{takayama\_kenshi@cyberagent.co.jp}
\affiliation{%
  \institution{CyberAgent}
  \city{Tokyo}
  \country{Japan}
}

\author{Takafumi Taketomi}
\email{taketomi\_takafumi@cyberagent.co.jp}

\affiliation{%
  \institution{CyberAgent}
  \city{Tokyo}
  \country{Japan}
}

\author{Takeshi Oishi}
\email{oishi@cvl.iis.u-tokyo.ac.jp}
\affiliation{%
  \institution{The University of Tokyo}
  \city{Tokyo}
  \country{Japan}
}


\begin{abstract}
We propose DeMapGS, a structured Gaussian Splatting framework that jointly optimizes deformable surfaces and  surface-attached 2D Gaussian splats.
By anchoring splats to a deformable template mesh, our method overcomes topological inconsistencies and enhances editing flexibility, addressing limitations of prior Gaussian Splatting methods that treat points independently.
The unified representation in our method supports extraction of high-fidelity diffuse, normal, and displacement maps, enabling the reconstructed mesh to inherit the photorealistic rendering quality of Gaussian Splatting.
To support robust optimization, we introduce a gradient diffusion strategy that propagates supervision across the surface, along with an alternating 2D/3D rendering scheme to handle concave regions.  
Experiments demonstrate that DeMapGS achieves state-of-the-art mesh reconstruction quality and enables downstream applications for Gaussian splats such as editing and cross-object manipulation through a shared parametric surface.
\end{abstract}

\begin{teaserfigure}
  \includegraphics[width=\textwidth]{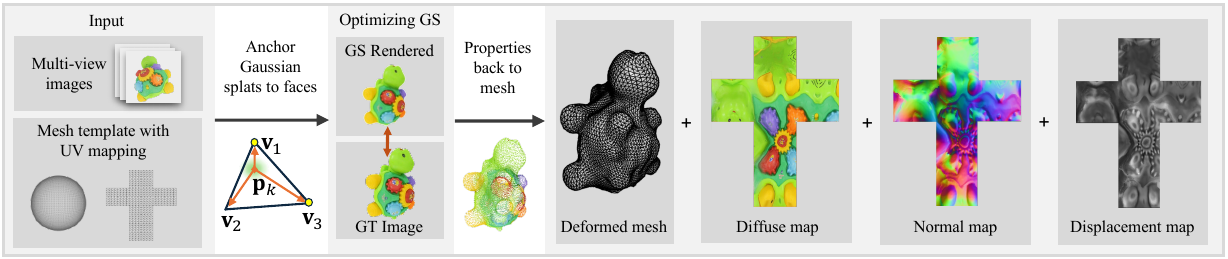}
  \caption{
  Overview of the DeMapGS framework. 
  Given multi-view images and an initial mesh template, we jointly optimize the vertex positions and surface-attached splats properties (Sec.~\ref{sec:parameterization}).  
  The optimization uses rendered outputs to compute losses, and incorporates our gradient diffusion mechanism for vertex updates (Sec.~\ref{sec:optimization}).  
  The optimized representation enables extraction of high-quality surface attribute maps (Sec.~\ref{sec:extraction}).
  Turtle model \textcopyright{} Google Research.
  }
  \Description{teaser}
  \label{fig:teaser}
  \vspace{0.5cm}
\end{teaserfigure}


\maketitle

\section{Introduction}
Reconstructing high-quality meshes from multiview images is essential for applications in computer vision, extended reality, and robotics~\cite{luo2024large}.
Recent advances in differentiable rendering, such as Neural Radiance Fields (NeRF)~\cite{mildenhall2021nerf} and 3D Gaussian Splatting (3DGS)~\cite{kerbl20233d}, have enabled high-fidelity image synthesis and have been increasingly adopted for shape reconstruction~\cite{azinovic2022neural, huang20242d}.
Despite these achievements, the resulting representations are often unstructured or implicit, which makes it difficult to extract accurate surface geometry or to enforce topological consistency.
Furthermore, meshes reconstructed from volumetric fields or splatting-based point clouds require dense meshing using marching cubes or other postprocesses~\cite{ma2024reconstructing, duisterhof2023deformgs}, which significantly increase their computational overhead while producing lower-quality surface meshes than the rendering outputs.

A growing body of work has explored attaching Gaussian splats to mesh surfaces to impose a geometric structure~\cite{shao2024splattingavatar, gao2024mani, choi2024meshgs, tobiasz2025meshsplats, guedon2024sugar}. 
This integration enables structured animation~\cite{rong2024gaussian, shao2024splattingavatar, liu2025dynamic}, editing~\cite{waczynska2024games}, and faster rendering~\cite{tobiasz2025meshsplats, guedon2024sugar}. 
However, with these approaches, the mesh is not directly optimized but serves only as a static anchor.
Other methods~\cite{ma2024reconstructing, li2024dgns, waczynska2024games} attempt to optimize the mesh geometry during training, but only support minimal deviations from the initial mesh structure.  

To simultaneously enable robust, large-step deformation of the mesh and the joint optimization of Gaussian parameters while ensuring that the resulting mesh satisfies the visual fidelity of Gaussian splatting, we propose a structured Gaussian splatting representation. 
The deformable surface preserves the mesh topology, thereby allowing the framework to naturally support animation, editing, and cross-object manipulation by maintaining structural and semantic consistency.  
One key novelty of our framework is the surface-aligned attribute extraction process, which distills the geometric and photometric surface attributes from the splats into high-quality diffuse, displacement, and normal maps.
The extracted explicit surface representation is compatible with standard graphics pipelines such as mipmapping, GPU tessellation, and multi-resolution rendering~\cite{luna2008introduction, shreiner2009opengl, sellers2016vulkan}, and also supports direct integration into content creation tools such as Blender or game engines for downstream editing and reuse.

Our method begins by introducing a parameterization scheme that anchors Gaussian splats onto mesh faces, enabling a unified framework for jointly optimizing splat parameters and mesh deformation.
To enable robust deformation optimization, we design a dedicated pipeline that integrates 2DGS–3DGS alternating rendering and a gradient diffusion strategy, which allows large-step vertex updates while ensuring smooth and regular mesh behavior.
After optimization, we extract high-quality surface attributes—diffuse, displacement, and normal maps—directly from the resulting splats through a dedicated postprocessing pipeline.

Our main contributions includes: 
\begin{enumerate}[itemsep=0pt, topsep=2pt]
    \item We propose an optimization framework that incorporates a gradient diffusion strategy, enabling large-step vertex updates while maintaining mesh regularity.
    \item We introduce a postprocessing pipeline that extracts high-fidelity surface attributes—diffuse, normal, and displacement maps—directly from the optimized Gaussian splats.
    \item We demonstrate that our structured Gaussian representation achieves high-quality geometry reconstruction and supports downstream applications such as editing, cross-object manipulation, and rendering.
\end{enumerate}

\section{Related Work}
In this section, we review key approaches to mesh deformation and recent advances in geometry reconstruction using 3DGS.

\subsection{Mesh Reconstruction from Multi-view Images}
Mesh-based representations have long been fundamental to geometry reconstruction, with methods for recovering 3D geometry from multi-view images evolving from traditional multi-view stereo~\cite{schoenberger2016mvs, shrestha2021meshmvs} to differentiable mesh rendering~\cite{kato2018neural, liu2019soft, shen2021deep, shen2023flexible}.
With the advent of NeRF~\cite{mildenhall2021nerf}, researchers have increasingly adopted differentiable volumetric representations as intermediate forms in the extraction of high-quality meshes~\cite{rematas2022urban, tang2023delicate, wei2024meshlrm, wei2025neumanifold}.
Several NeRF-based extensions further incorporate signed distance functions to enable the extraction of smooth and continuous surfaces~\cite{wang2021neus, munkberg2022extracting, azinovic2022neural, li2023neuralangelo, wang2023neus2, cao2024supernormal}.

3DGS~\cite{kerbl20233d} represents scenes by placing anisotropic Gaussians in 3D space and optimizing their spatial and appearance parameters for novel view synthesis, and it achieves real-time performance with high-quality results. 
Recent methods for recovering a mesh from Gaussians typically perform mesh reconstruction using post-processing techniques—such as depth-based fusion~\cite{wolf2024gs2mesh, huang20242d, yu2024gaussian, turkulainen2024dn}, multi-view normal-guided surface extraction~\cite{wang2024gaussurf, dai2024gaussian}, or Poisson surface reconstruction~\cite{guedon2024sugar, zhao2024dygasr}—using rendered depth maps and estimated normal maps.

\subsection{Surface Deformation}
Deformable surface models aim to represent 3D geometry while preserving topological consistency and enabling stable UV mappings; these are key properties for applications such as tracking, texture mapping, and animation~\cite{montagnat2001review}.
In this section, we focus specifically on mesh-based deformation.

Many approaches introduce data-driven external forces that act as attractors and guide the surface toward salient features such as edges, contours, or keypoints~\cite{cohen1993balloon, kass1988snakes, li2005bvf}.
Internal energy terms—such as elastic, membrane, or thin-plate energies—help maintain local smoothness and prevent unrealistic distortions~\cite{terzopoulos1987elastically, bookstein1989principal, sorkine2007rigid, amberg2007optimal}.
Recent methods have integrated these priors into differentiable models: Pixel2Mesh++~\cite{wen2019pixel2mesh++} and MeshMVS~\cite{shrestha2021meshmvs} deform meshes using graph-based operations.
Nicolet et al.~\cite{nicolet2021large} introduced a regularization strategy that minimizes Laplacian energy by diffusing first-order gradients, thereby enabling stable large-step updates without sacrificing geometric plausibility.

\subsection{Mesh-anchored Gaussian Splatting} 
To address the lack of structure and topological meaning in GS, recent methods have anchored splats directly onto mesh surfaces using barycentric coordinates or local offsets.
Several works have adopted this strategy in animation \rev{or 4D reconstruction~\cite{shao2024splattingavatar, rong2024gaussian, luo2024mags, ma2024mags, wang2024canfields}}, for editing and manipulation~\cite{waczynska2024games, gao2024mani, guedon2024frosting}, or to facilitate efficient 3D reconstruction and high-quality rendering~\cite{guedon2024sugar, choi2024meshgs, scolari2024mesh2splat}.
With these approaches, the mesh is either fixed or is allowed only limited deformation. As a result, these approaches depend heavily on accurate mesh initialization and are not applicable to general reconstruction from arbitrary templates.


In contrast, our method introduces a structured Gaussian splatting framework that enables the large-step deformation of a mesh template while preserving the fine-grained geometry and appearance of the target object. 
By extracting diffuse, displacement, and normal maps from the optimized splats, our approach bridges the gap between splatting-based rendering and structured mesh representations that are suitable for downstream graphics applications.

\section{Problem Definition}
\label{sec:problem_definition}
We address two key limitations in 3D scene modeling: (1) the unstructured nature of Gaussian splatting, which lacks topological coherence and limits feature transfer or editing; and (2) the challenge of deforming template meshes to match an observed geometry while maintaining high rendering fidelity. 

To this end, we propose a unified representation that integrates deformable surface modeling with structured Gaussian splatting. 
Starting with a template mesh defined by vertices $\smash{\mathcal{V} = \{\bm{v}_i\}_{i=1}^{N^{\mathcal{V}}}}$ and faces $\smash{\mathcal{F} = \{\bm{f}_j\}_{j=1}^{N^{\mathcal{F}}}}$, our method jointly optimizes mesh deformation and a set of surface-attached 3D or 2D Gaussian splats $\smash{\mathcal{G} = \{\bm{g}_k\}_{k=1}^{N^{\mathcal{G}}}}$ 
from a set of multi-view images $\smash{\mathcal{I} = \{I_l\}_{l=1}^{N^{\mathcal{I}}}}$:
\begin{equation}  
\begin{aligned}
    (\hat{\mathcal{V}}, \hat{\mathcal{G}}) &=\arg\min_{\mathcal{V},\,\mathcal{G}}\mathcal{L}(\mathcal{V},\mathcal{F},\mathcal{G}, \mathcal{I}, \Pi),\\
    \text{where }\mathcal{L} &= \mathcal{L}_{\text{photo}} + \mathcal{L}_{\text{ssim}}  + \mathcal{L}_\text{reg} + \mathcal{L}_{\text{normal}} + \mathcal{L}_{\text{dist}},
\end{aligned}
\end{equation}
%
%
where $\smash{\Pi=\{\pi_l\}_{l=1}^{N^{\mathcal{I}}}}$ contains the known camera projection functions. 
The numbers of vertices $\smash{N^{\mathcal{V}}}$ and faces $\smash{N^{\mathcal{F}}}$ are fixed, whereas the number of splats $\smash{N^{\mathcal{G}}}$ varies according to the adaptive strategy of 3DGS~\cite{kerbl20233d}. 
Each Gaussian follows the 3DGS formulation
and is parameterized by its center $\bm{p}_k \in \mathbb{R}^3$, rotation $\bm{q}_k$ (a quaternion in scalar-first format), scale $\bm{s}_k \in \mathbb{R}^3$, opacity $o_k \in \mathbb{R}$, and color $\bm{c}_k \in \mathbb{R}^3$: $\bm{g}_k = (\bm{p}_k, \bm{q}_k, \bm{s}_k, o_k, \bm{c}_k)$. 
During our alternative optimization (Sec. \ref{sec:optimization-flow}), we maintain the same parameters for 3DGS and 2DGS~\cite{huang20242d}, whereas we omit the third component of $\bm{s}_k$ in the 2DGS framework. 



We render the Gaussian splats and apply the photometric loss $\mathcal{L}_{\text{photo}}$ and the SSIM loss $\mathcal{L}_{\text{ssim}}$ to align the rendered and input colors. 
To secure stable convergence under 2D supervision, we add the bi-Laplacian regularization $\mathcal{L}_\text{reg}$ and adopt a large-step gradient update inspired by~\cite{nicolet2021large}. 
Our optimization merges 3DGS and 2DGS rasterization 
and incorporates the normal consistency loss $\mathcal{L}_{\text{normal}}$ and depth distortion loss $\mathcal{L}_{\text{dist}}$ from 2DGS.
While $\mathcal{L}_{\text{normal}}$ is typically computed by comparing splat normals to normals estimated from the rendered depth map,  
our framework also supports stronger supervision when reference or ground-truth normal maps are available.  
In such cases, $\mathcal{L}_{\text{normal}}$ can instead be evaluated against these external normals, enabling more accurate recovery of fine surface details.

One of the key contributions of our method is its ability to extract high-resolution diffuse, displacement, and normal maps ($\mathcal{T}$, $\mathcal{D}$, and $\mathcal{N}$) from the optimized model:
\begin{equation}
    \Gamma: (\hat{\mathcal{V}}, \mathcal{F}, \hat{\mathcal{G}}) \rightarrow (\mathcal{T}, \mathcal{D}, \mathcal{N}).
\end{equation}
%
The attribute extraction procedure is detailed in Sec.~\ref{sec:extraction}. 
As described in Sec.~\ref{sec:gs_attach}, each splat holds displacement from its corresponding face and rotation relative to it; this arrangement allows the model to represent fine geometry as well as color.
That is, the displacement map enables subpixel geometric detail via GPU tessellation, whereas normal and diffuse maps enhance photorealistic rendering quality. 

\begin{figure}[t]
    \centering
    \includegraphics[width=\linewidth]{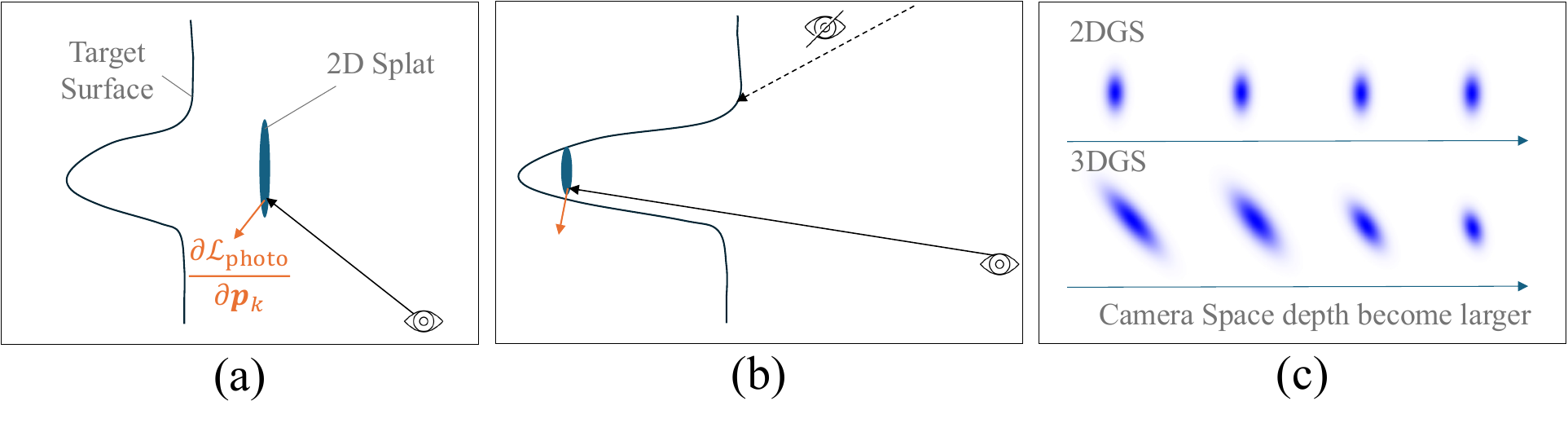}
    \vspace{-0.4cm}
    \caption{
    (a) Gradients of 2D splat positions lie in the plane perpendicular to the view direction;
    (b) In concave regions, splats are only visible from frontal views, yielding weak gradients along the surface normal.
    \rev{(c) Illustration of variation in projected 2D covariance for 2DGS and 3DGS under changes in the camera-space depth of the Gaussian mean.}
    }
    \label{fig:3d-2d}
\end{figure}
\section{Gaussian Splats on Deformable Surface}
\label{sec:parameterization}

This section presents our structured Gaussian splatting representation defined on a deformable mesh surface.  
We first describe how Gaussian splats are attached to a template mesh and jointly optimized with both global alignment and vertex-level deformation.  
We then detail how each splat is parameterized relative to its associated triangle using barycentric coordinates, normal displacement, and local rotation.  
To ensure splats remain consistently attached to the surface during deformation, we introduce a lightweight scheme to handle walk-on-triangle cases.
\begin{figure*}[t]
    \centering
    \includegraphics[width=\linewidth]{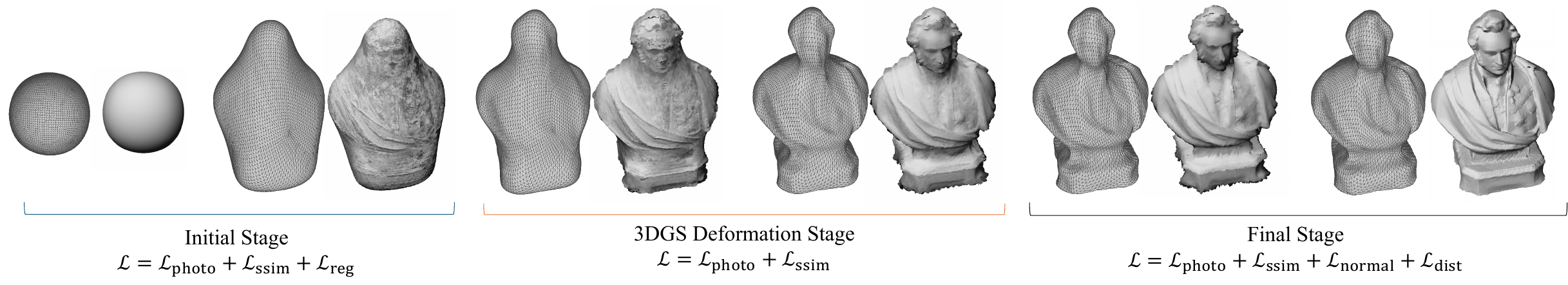}
    \caption{Detail optimization stages.  
    Each stage employs different loss terms tailored to its purpose, as shown.}
    \Description{Training stages.}
    \label{fig:iteration}
\end{figure*}
\subsection{Deformable Surface with GS}
We introduce a deformable mesh-anchored Gaussian splatting approach. 
First, we attach a set of Gaussian splats uniformly to a template mesh that covers the target object. 
Next, we deform the mesh by moving vertices $\mathcal{V}$ and optimizes the attached splats $\mathcal{G}$ using the differentiable rendering and backpropagation framework. 
We use 2DGS~\cite{huang20242d} and 3DGS~\cite{kerbl20233d} as model representations and differentiable renderers during the optimization. 



In addition to the per-vertex deformations, we simultaneously optimize a global rigid transformation of the mesh to enhance its robustness to variations in the initial mesh template;  
the optimization becomes less sensitive to differences in the scale, rotation, or position of the template mesh relative to the target object.
%
The global transformation is composed of a scaling vector $\bm{s}^\ast \in \mathbb{R}^3$, a rotation $\bm{q}^\ast$, and a translation vector $\bm{t}^\ast \in \mathbb{R}^3$. 
The globally transformed vertex position is given as:
\begin{equation}
    \bm{v}' = \bm{S}^\ast R(\bm{q}^\ast) \bm{v} + \bm{t}^\ast,
\end{equation}
where $\bm{S}^\ast$ is the diagonal matrix formed from $\bm{s}^\ast$, and $R(\bm{q}^\ast)$ denotes the rotation matrix converted from the quaternion $\bm{q}^\ast$. 
We simplified $\bm{S}^\ast R(\bm{q}^\ast)$ as $M$: $M = \bm{S}^\ast R(\bm{q}^\ast)$.  

Throughout the optimization, we adopt adaptive densification, splitting, and opacity reset strategies as was done in~\cite{kerbl20233d}; these enhance convergence and improve rendering fidelity.
Although the splats are constrained to lie on the deformable mesh surface, these strategies preserve the expressive power of unstructured Gaussian splatting and enable high-quality photorealistic reconstruction.

\subsection{GS Representation on Mesh Surface}
\label{sec:gs_attach}
Each Gaussian splat $\bm{g}_k$ is attached to a specific face $\bm{f}_{<k>}$ on the mesh surface.
Its center position $\bm{p}_k$ is represented using barycentric 
coordinates $\boldsymbol{\beta}_k = \{\beta_{k,m}\}_{m=1}^3$ over the globally transformed face vertices $\{\bm{v}'_{\bm{f}_{<k>},m}\}_{m=1}^3$, with displacement $d_k \in \mathbb{R}$ along the face normal $\bm{n}_{\bm{f}_{<k>}}$:
\begin{equation}
    \bm{p}_k = \sum_{m=1}^3 \beta_{k,m} \bm{v}'_{\bm{f}_{<k>},m} + d_k \bm{n}_{\bm{f}_{<k>}}. 
\end{equation}
%
Here, $\bm{\beta}_k$ and $d_k$ are optimizable parameters that are updated through the optimization process together with the vertex positions. 
The face normal $\bm{n}_{\bm{f}_{<k>}}$ is recomputed each iteration to reflect updates of the vertices. 
The optimized displacement values derive the displacement map $\mathcal{D}$. 

We represent the rotation $\bm{q}_k$ of each splat in the world space using the coordinate frame for its corresponding face. 
This strategy enables stable convergence and a direct interpretation of the per-splat rotation as components of the normal map $\mathcal{N}$. 
We use the local rotation $\bar{\bm{q}}_k$ with respect to the face it belongs to with a face-based rotation $\mathbf{q}^f_{<k>}$ to yield $\bm{q}_k$, as follows: 
\begin{equation}
       \bm{q}_k = \bar{\bm{q}}_k \otimes\mathbf{q}^f_{<k>},
       \label{eq:normal-rotation}
\end{equation}
where $\otimes$ denotes the Hamilton product between the two quaternions. 
During the optimization process, $\bar{\bm{q}}_k$ is updated instead of being directly updating $\bm{q}_k$. 
The face-based rotation $\smash{\mathbf{q}^f_{<k>}}$ aligns the canonical face normal $(0, 0, 1)$ with $\bm{n}_{\bm{f}_{<k>}}$ and is computed as:
%
\begin{align}
    \mathbf{q}^f_{<k>} &= \frac{\hat{\mathbf{q}}^f_{<k>}}{\|\hat{\mathbf{q}}^f_{<k>}\|}, \\
    \text{where } \hat{\mathbf{q}}^f_{<k>} &=\left(1 + \bm{n}_{f_{<k>}[3]},\ -\bm{n}_{f_{<k>}[2]},\ \bm{n}_{f_{<k>}[1]},\ 0\right).
\end{align}
%
Here, $\bm{n}_{[x]}$ denotes the $x$-th component of the normal vector.
We also jointly optimize the per-splat scale $\bm{s}_k$, opacity $o_k$, and color $\bm{c}_k$ according to the 3DGS framework~\cite{kerbl20233d}. 
The latter two, $o_k$, and $\bm{c}_k$, together serve as a diffuse map $\mathcal{T}$ in our representation.

\subsection{Handling Walk-on-Triangles}
Ideally, splats move only along their surface normal during optimization, but Gaussian splatting gradients often cause tangential drift~\cite{zhou2025robust}.
This can push splats out of their assigned triangles, known as {\it walk-on-triangles}.
SplattingAvatar~\cite{shao2024splattingavatar} adds an MSE loss that penalizes the distance between each splat and its projection onto the surface of the mesh. 
Although this is effective, it incurs additional computational cost and can still drive splats toward incorrect positions.

To address this problem, we employ a lightweight three-step update strategy considering the fact that a barycentric coordinate turns negative exactly when the splat crosses the edge that is opposite the corresponding vertex:
\begin{enumerate}[itemsep=0pt, topsep=2pt]
\item 
Clamp any negative barycentric coordinate to zero.
\item 
Renormalize the remaining coordinates so that they sum to one.
\item 
Reassign the splat to the neighboring triangle that lies across the edge opposite the clamped vertex, as defined by the nonzero barycentric weights.
\end{enumerate}
%
%
Note that if only one coordinate remains positive, the splat is assigned to any adjacent triangle across a negative edge, and then moved across neighbors until the correct face is reached.


\section{Optimization}
\label{sec:optimization}
Beyond the standard loss terms ($\mathcal{L}_{\text{photo}}$, $\mathcal{L}_{\text{ssim}}$, $\mathcal{L}_{\text{reg}}$) described in Sec.~\ref{sec:problem_definition}, our method addresses two key challenges unique to deformable surface splatting.  
First, we alternate between 2DGS and 3DGS rendering to benefit from both accurate surface supervision (e.g., for normal maps) and strong volumetric gradients for stable geometry optimization.  
Second, we propose a gradient diffusion strategy that propagates per-splat gradients to the surrounding mesh, enabling robust and coordinated vertex updates.
\label{sec:shape}

\subsection{Alternative 2D-3DGS Optimization}
\label{sec:optimization-flow}

To extract a high-quality normal map, we use the 2DGS framework \cite{huang20242d}, which represents each splat as an elliptical plate aligned with the surface.

\noindent \textbf{2DGS for Surface-Aligned Optimization: }
%
2DGS places each splat as a thin, 2D ellipse in 3D space. 
The third column of $R({\bm{q}_k})$ gives the normal of splat surface $\bm{n}_k^\mathcal{G}$, which matches  Eq.~\ref{eq:normal-rotation} by:
\begin{equation}
    \bm{n}_k^\mathcal{G} = R({\bar{\bm{q}}_k}) \bm{n}_{f_{<k>}}.
\end{equation}
Since each splat is treated as a true 2D surface, we can compute accurate ray-splat intersections, which enables precise rendering of per-pixel depth and normal maps.
This capability allows us to incorporate two regularization terms from 2DGS:
(1) {Normal consistency loss} $\mathcal{L}_{\text{normal}}$ to match  $\bm{n}_k^\mathcal{G}$ with normals from the depth map (or other reference normal maps);
(2) {Depth distortion loss} $\mathcal{L}_{\text{dist}}$ to keep ray-splat intersections tightly clustered in depth along each viewing ray.

Another benefit of 2DGS is its surface-aware gradient: forces shift the mesh face that anchors each splat rather than the splat’s thickness as in 3DGS. 
This makes 2DGS ideal for the early phase when we need robust, large-step vertex moves for coarse alignment.



\noindent \textbf{Alternating Optimization Schedule: }
As noted by \cite{zhou2025robust}, positional gradients for 2D splats are constrained to lie within the plane perpendicular to the view direction.
When a splat lies inside a concave region, it becomes visible only when the view direction is nearly orthogonal to the splat plane (Fig. ~\ref{fig:3d-2d})—yet in such cases, the positional gradients lie almost entirely within that plane, lacking the vertical component needed to move the splat deeper into the concavity. 

\rev{
In contrast, 3DGS offers full-directional positional gradients. 
As illustrated in Fig.~\ref{fig:3d-2d}(c), translating a 3D splat along the camera ray alters its projected 2D covariance due to its inhomogeneous dependence on camera-space inverse depth. 
This depth-sensitive behavior prevents the gradient from collapsing onto any fixed plane. 
A formal proof is provided in the Appendix.}

Our pipeline runs in three stages as in Fig.~\ref{fig:iteration}:
\begin{enumerate}[itemsep=0pt, topsep=2pt]
    \item 2DGS for stable, large-scale deformation.
    \item 3DGS for detailed volumetric sculpting.
    \item 2DGS for final surface consistency.
\end{enumerate}
In the last stage, we apply the normal and depth distortion losses from~\cite{huang20242d} to achieve a physically plausible finish.

\begin{table*}[t]
    \centering
    \small
    \caption{Geometry evaluation. CD1 and CD2 represent one-sided Chamfer distances: CD1 measures the average squared distance from predicted points to ground truth, while CD2 measures the reverse. Reported CD values are scaled by $\times 10^{-4}$, and SD by $\times 10^{-2}$. Lower is better.}
    \begin{tabular}{|c|c @{\hspace{4pt}} c @{\hspace{4pt}} c|c @{\hspace{4pt}} c @{\hspace{4pt}} c|c @{\hspace{4pt}} c @{\hspace{4pt}} c|c @{\hspace{4pt}} c @{\hspace{4pt}} c|c @{\hspace{4pt}} c @{\hspace{4pt}} c|c @{\hspace{4pt}} c @{\hspace{4pt}} c|}
    \hline
        & \multicolumn{3}{|c|}{Tree} & \multicolumn{3}{c|}{Bust} & \multicolumn{3}{c|}{Buddha} & \multicolumn{3}{c|}{Book} & \multicolumn{3}{c|}{Box} & \multicolumn{3}{c|}{Cat}\\\cline{2-19}
         & CD1 & CD2 & SD & CD1 & CD2 & SD & CD1 & CD2 & SD & CD1 & CD2 & SD & CD1 & CD2 & SD & CD1 & CD2 & SD \\\hline
        SuGaR & \underline{0.13}& \underline{0.12}&0.69 & 1.27& 1.66& 1.96& \underline{0.13}& 0.18& 1.14& \underline{0.24}& 0.36& 1.24& 2.82& 6.26& 3.26& 1.93 & 24.43 & 6.09\\
       2DGS  & 0.14& \underline{0.12}& \underline{0.65}& 1.37& 1.14& \underline{1.33}& \textbf{0.10} & \textbf{0.09} & \textbf{0.62} & \textbf{0.20} & \underline{0.16} & \textbf{0.65} & 1.04& \underline{1.17}& \underline{1.28}& \textbf{0.20} & \textbf{0.20} & \textbf{1.36} \\
       GOF & 0.22& 1.93& 2.41 & \underline{0.53}& \underline{0.90} & 1.58 & \underline{0.13}& 2.48 & 2.08 & 0.67 & 6.29 & 2.08 & 4.10 & 9.57 & 4.76 & \underline{0.21} & 2.21 & 2.39\\
       DG-mesh & 0.17 & 0.24 & 0.69 & 0.98 & 1.05 & 1.32 &0.31 &14.98 &4.92 & 0.42&0.43&0.86&\underline{0.31}&14.90&4.91&0.29&\underline{0.23}&\underline{1.39}\\
       NRICP &2.52 &0.76 &1.65 & 59.00&4.11&11.00&11.03 &1.96 &3.94 & 7.39&0.76 &3.37 &3.07 &1.29 &2.11 & 7.32 &1.03 &3.49 \\
       Point2Mesh &0.44 &1.01 &8.05 & 0.90&3.85 &6.95 &0.67 &4.24 &5.62 & 1.30&3.41 &5.00 &0.81 &5.05 &8.43 &1.05 &5.53 &5.59 \\
       GaMeS  &22.11&69.57&14.18& 29.14&15.41&16.90&36.03&44.26&17.84& 18.11&31.38&11.52&69.99&65.74&21.83&38.00&44.87&18.56\\
       Ours-s & \textbf{0.11}& \textbf{0.11}& \textbf{0.62}& \textbf{0.42}& \textbf{0.35}& \textbf{1.21}& \underline{0.13}& \underline{0.12}& \underline{0.72}& 0.25& \textbf{0.15} & \underline{0.67}& \textbf{0.15} & \textbf{0.14} & \textbf{0.71}& 0.45& 0.55& 1.67\\
       
       \hline
    \end{tabular}
    \label{tab:geom}
\end{table*}
\begin{table*}[t]
    \centering
    \small
    \caption{Mesh texture quality and rendering speed. 
    PSNR (↑) and LPIPS (↓) evaluate rendering quality, while FPS (↑) measures rendering efficiency. 
    LPIPS is reported in $\times 10^{-5}$.
    Since GOF and DG-Mesh do not provide texture extraction pipelines, their PSNR and LPIPS are not reported.}
    \begin{tabular}{|c|c @{\hspace{4pt}} c @{\hspace{4pt}} c|c @{\hspace{4pt}} c @{\hspace{4pt}} c|c @{\hspace{4pt}} c @{\hspace{4pt}} c|c @{\hspace{4pt}} c @{\hspace{4pt}} c|c @{\hspace{4pt}} c @{\hspace{4pt}} c|c @{\hspace{4pt}} c @{\hspace{4pt}} c|}
    \hline
        & \multicolumn{3}{|c|}{Tree} & \multicolumn{3}{c|}{Bust} & \multicolumn{3}{c|}{Buddha} & \multicolumn{3}{c|}{Book} & \multicolumn{3}{c|}{Box} & \multicolumn{3}{c|}{Cat}\\\cline{2-19}
         & PSNR & LPIPS & FPS 
         & PSNR & LPIPS & FPS 
         & PSNR & LPIPS & FPS 
         & PSNR & LPIPS & FPS 
         & PSNR & LPIPS & FPS 
         & PSNR & LPIPS & FPS \\\hline
        SuGaR &32.60&0.66&9.72& 33.72&0.48&10.56&31.16&\textbf{0.55}&10.24& \textbf{33.99}&\textbf{0.28}&10.09&31.64&\underline{0.44}&9.46&21.40 &2.69 &9.15\\
       2DGS  &24.13&2.12&\underline{32.19}& 22.99&2.09&34.79&19.11 &6.44 &\underline{32.46}& 25.82&1.17&\underline{40.28}&29.56&0.59&\underline{36.27}&24.58&1.61&\textbf{49.53}\\
       GOF &- &- &3.25& - &- &9.40&- &- &4.64& - &- &6.33&- &-&4.96&- &- &10.30\\
       DG-mesh &- &- &25.43& -&- &\underline{37.06}&- &- &32.21& - &- &25.11&- &- &16.39&- &- &32.08\\
       Ours-s&\textbf{33.42}&\underline{0.61} &\textbf{43.38} & \underline{35.33}&\underline{0.23}&\textbf{47.44}&\textbf{31.17}&0.64&\textbf{45.02} & \underline{33.36}&\underline{0.36}&\textbf{48.16}&\textbf{31.99}&\textbf{0.40}&\textbf{48.05} &\textbf{31.83}&\underline{0.46}&\underline{47.52} \\
       Ours-d & \underline{33.31}& \textbf{0.60}& 16.75 & \textbf{35.37}& \textbf{0.21}& 16.72 & \underline{31.16}& \underline{0.63}& 16.77 & 33.07 & 0.37 & 17.15 & \underline{31.95}& \textbf{0.40}& 17.12 & \underline{31.77}& \textbf{0.45}& 16.78
       \\
       
       \hline
       
    \end{tabular}

    \label{tab:mesh}
\end{table*}
\subsection{Gradient Diffusion}
\label{sec:gradient}


Our mesh-guided framework directs gradients to the fixed-topology mesh vertices in Cartesian space as follows:
%
%
\begin{align}
    \bm{v}_i &\leftarrow \bm{v}_i + \Delta \bm{v}_i + \Delta\bm{m}_i, \\
    \Delta \bm{v}_i &= - \eta \left[\frac{\partial \mathcal{L}}{\partial\bm{V}}\right]_{[i]}, 
 \text{where } 
\left[\frac{\partial \mathcal{L}}{\partial\bm{V}}\right]_{[i]}=
    M^{-1} \sum_{\bm{g}_k \in \mathcal{G}_{<\bm{v}_i>}} \beta_{k,i} \frac{\partial \mathcal{L}}{\partial \bm{p}_k}.
\end{align}
 $\mathcal{G}_{<\bm{v}_i>}$ denotes the set of splats on the faces adjacent to $\bm{v}_i$, $\beta_{k, i}$ is the barycentric weight of $\bm{g}_k$ for vertex $\bm{v}_i$,
 $\left[ \cdot \right]_{[i]}$ denotes the $i$-th row element of $\left[ \cdot \right]$, and $\eta$ is the learning rate.
The term $\Delta\bm{m}_i$ represents the momentum update, whose calculation depends on optimizer (e.g., from Adam~\cite{kingma2014adam}).
%
To propagate localized splat gradients across the mesh and achieve a smooth deformation, we adopt a large-step gradient diffusion strategy~\cite{nicolet2021large} in our framework.

\subsubsection{Splat-aware Vertex Gradient Diffusion}
Instead of directly optimizing the vertex positions $\bm{v}$ using raw gradients, the method first applies a diffusion-like smoothing step, analogous to heat diffusion, prior to optimization. 
The vertex update is given by:
\begin{equation}
\Delta \bm{v}_i^\ast = - \eta \left[(\bm{I} + \lambda_l \bm{L})^{-2} \frac{\partial \mathcal{L}}{\partial\bm{V}}\right]_{[i]},
\label{eq:large-step}
\end{equation}
where $\bm{I}$ is the identity matrix, $\bm{L}$ is the Laplacian matrix of the mesh, and $\lambda_l$ is a hyperparameter that controls the strength of the regularization.

\begin{figure}
    \centering
    \includegraphics[width=\linewidth]{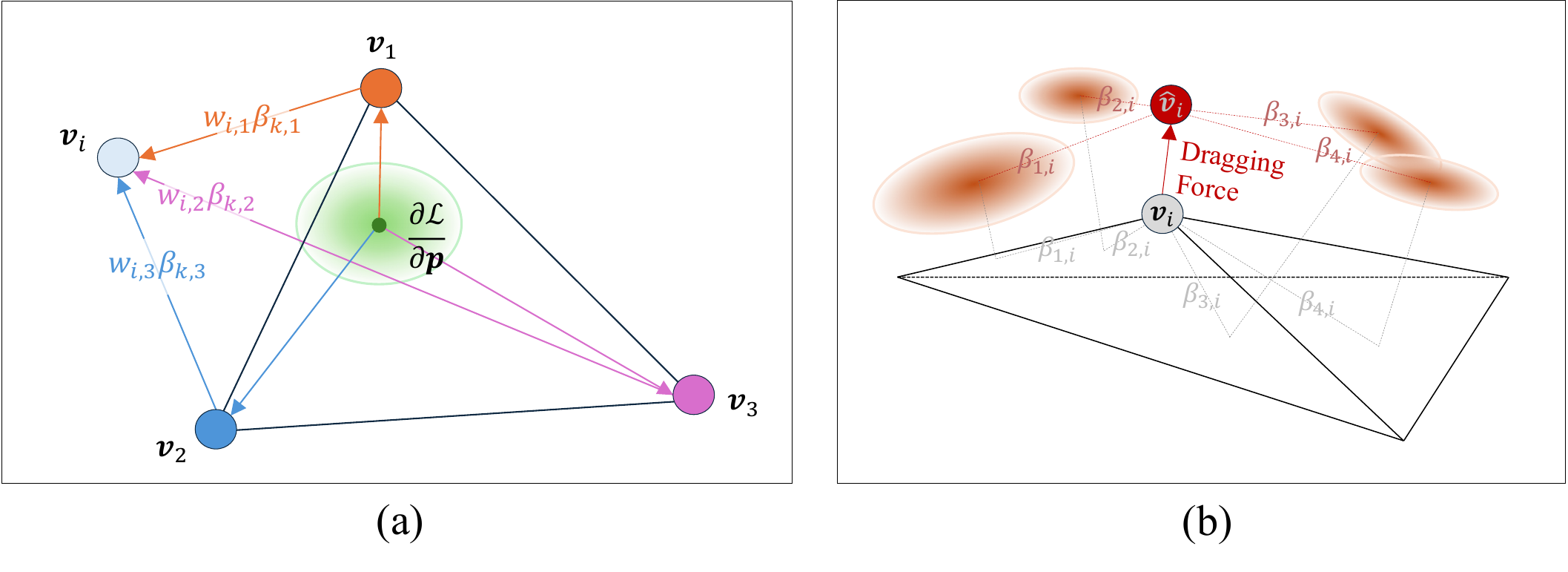}
    \vspace{-0.4cm}
    \caption{(a) Gradient flow from splats to vertex $\bm{v}_i$ (omitting $M$): gradients are first propagated to vertices on the attached face, then diffused to $\bm{v}_i$ via weighted averaging.
    (b) Vertex realignment: target position is computed as a weighted average of neighboring splat centers.}
    \Description{Illustration of gradient diffusion}
    \label{fig:gradient}
\end{figure}

In our framework, this induces a hierarchical update process in which each vertex $\bm{v}_i$ aggregates gradients from all other vertices $\bm{v}_{l}$ as shown in Fig.~\ref{fig:gradient}a:
%
%
\begin{align}
\Delta \bm{v}_i^\ast &= - \eta 
M^{-1} \sum_{l=1}^{N^{\mathcal{V}}}  w_{i,l} \left( \sum_{\bm{g}_k \in \mathcal{G}_{<\bm{v}_l>}} \beta_{k, l} \frac{\partial \mathcal{L}}{\partial \bm{p}_k} \right),\\
%
%
%
w_{i,l} &= \left[ (\bm{I} + \lambda_l \bm{L})^{-2} \right]_{[i,l]},
\end{align}
%
%
where $w_{i,l}$ is the diffusion weights, which naturally decays with an increase in the geodesic distance at the mesh surface.
$\left[ \cdot \right]_{[i,l]}$ denotes the $i$-th row and $l$-th column element of $\left[ \cdot \right]$.  


\subsubsection{Vertex Re-Alignment by Displacement of Splats}
Although the above-described approach successfully smooths the vertex updates, it does not directly regularize the displacement parameters $d_k$. 
As a result, if displacements are updated aggressively, the corresponding vertices may lag behind, which leads to inconsistencies between the mesh geometry and splat positions.
A naive solution is to reset the displacement every several iterations.
However, using this alone slows down the entire optimization process.

To mitigate the mismatch between the vertex positions and the displaced splats, we introduce a realignment step that pulls the mesh vertices toward the updated splat centers.
We compute a target position (Fig.~\ref{fig:gradient}b) for each vertex $\hat{\bm{v}}_i$ as a weighted average of the splat centers as:
\begin{equation}
\hat{\bm{v}}_i = \frac{ \sum_{\bm{g}_k \in \mathcal{G}_{<\bm{v}_i>}} \beta_{k, i} \bm{p}_k }{ \sum_{\bm{g}_k \in \mathcal{G}_{<\bm{v}_i>}} \beta_{k, i} }.
\label{eq:vertex-realign}
\end{equation}
%
We then diffuse the dragging force to nearby vertices and update the vertices using:
\begin{equation}
\Delta \bm{v}_i^{+} = 
M^{-1}\left[(\bm{I} + \lambda_l \bm{L})^{-2} (\hat{\bm{V}} - \bm{V})\right]_{[i]}, 
\label{eq:vertex-update}
\end{equation}
where $\hat{\bm{V}} = [\hat{\bm{v}}_1, \hat{\bm{v}}_2 ..., \hat{\bm{v}}_{N^{\mathcal{V}}}]^{\text{T}}$. 
This realignment is performed every 50 iterations with negligible additional computational cost.
This step ensures that the vertex positions remain synchronized with the evolving splat geometry, particularly in regions having strong curvature.


\section{Surface Attribute Extraction}
\label{sec:extraction}
After optimization, we extract the diffuse, normal, and displacement maps from the learned splats by treating each triangle face of the deformed mesh as a local 2D image plane. 
\begin{figure*}[t]
    \centering
    \includegraphics[width=\linewidth]{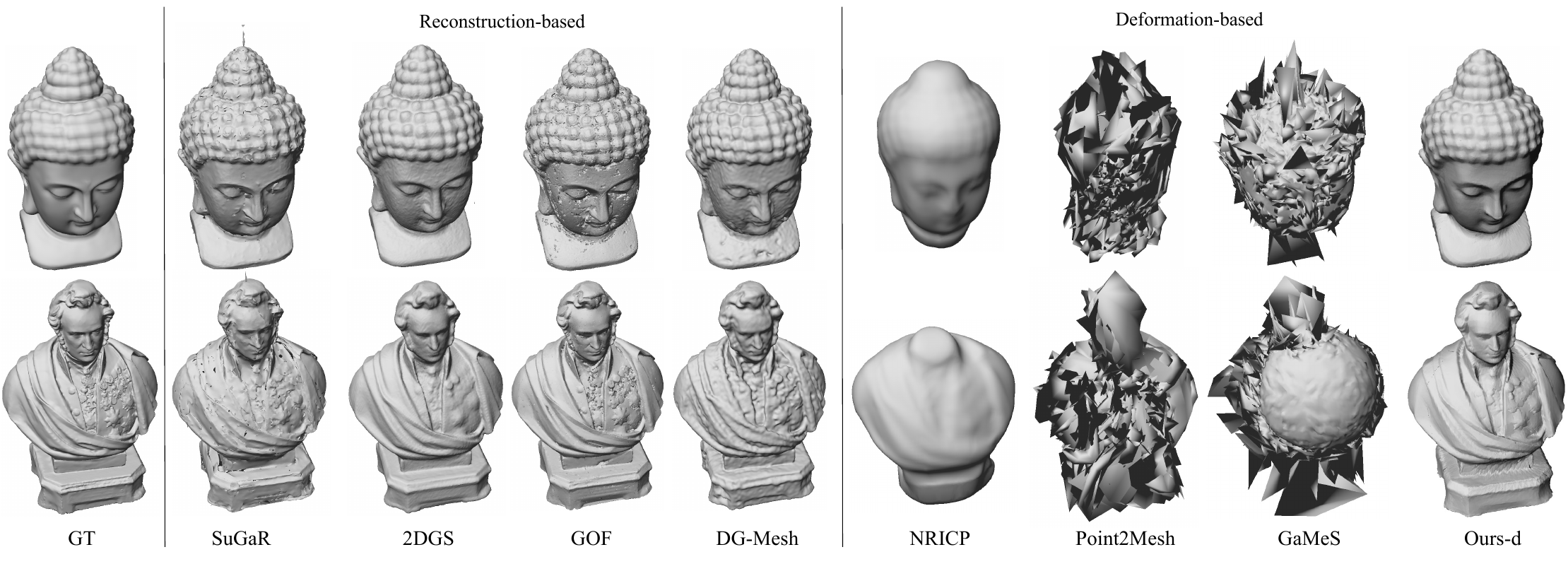}
    \caption{OpenGL-rendered geometry of reconstructed meshes. Deformation-based methods fail to fully deform to the target shape or produce noisy geometry.
    Reconstruction-based methods tend to preserve the overall structure but may exhibit noise or distortion. 
    Our method, based on 2DGS rendering, closely resembles the 2DGS baseline. 
    Due to tessellation and normal maps, our results exhibit a smoother appearance.
    }
    \Description{Geometry result}
    \label{fig:geometry}
\end{figure*}
\begin{figure*}[t]
    \centering
    \includegraphics[width=\linewidth]{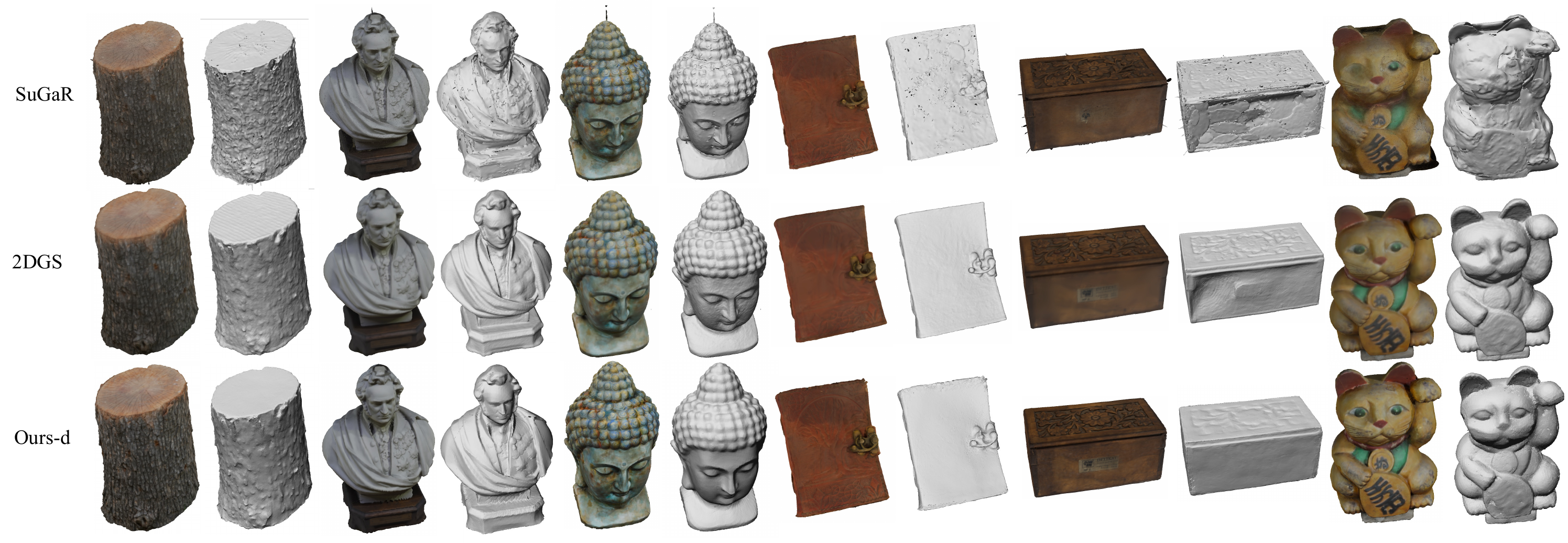}
    \caption{
    OpenGL-rendered results with and without texture.
    2DGS relies on per-vertex colors, resulting in visibly blurred appearance.  
    SuGaR uses diffuse maps, which lead to visually appealing results. However, the underlying geometry suffers from artifacts, resulting in unstable performance. 
    Our method achieves overall cleaner and sharper renderings by combining structured geometry with high-quality diffuse map extraction.
    Models \textcopyright{} matousekfoto (Tree), Geoffrey Marchal (Bust), hullo (Buddha), pigfinite (Book), 3D Master (Box), Barbucha Studio (Cat).
    }
    \label{fig:rgb-res}
\end{figure*}
\subsection{Projection Model:} 
Each triangle face $\bm{f}_j$ is treated as a local orthographic projection surface, analogous to an image plane in screen-space rendering. 
For an arbitrary texel $\bm{t}_{(j)}$ on $\bm{f}_j$, specified by barycentric coordinates $\hat{\boldsymbol{\beta}}_{(j)} \in \mathbb{R}^3$, we compute its 3D position in world space as:
\[
\hat{\bm{p}}_{(j)} = \sum^3_{m=1}\hat{\beta}_{(j),m} \bm{v}'_{j, m},
\]
where $\bm{v}'_{j,m}$ is the globally transformed vertex of face $\bm{f}_j$.

To evaluate the contribution of a splat $\bm{g}_k$ to the texel at $\hat{\bm{p}}_{(j)}$, 
we cast a ray along the face normal $\bm{n}_j$ from $\hat{\bm{p}}_{(j)}$, 
and compute the intersection of the ray with the splat's plane.
The 2D local coordinate $\hat{\bm{x}}_{(j),k} \in \mathbb{R}^2$ of the intersection point is computed as:
\begin{equation}
\begin{aligned}
 \hat{\bm{x}}_{(j),k} &= \left[\hat{\bm{p}}_{(j)} + \delta \bm{n}_j
 - \bm{p}_k\right]^{\text{T}}\cdot \bm{l}_k, \\ 
 &\text{where }\delta = \frac{(\bm{p}_k - \hat{\bm{p}}_{(j)})\cdot \bm{n}_k^{\mathcal{G}}}{\bm{n}_j \cdot \bm{n}_k^{\mathcal{G}}}.
\end{aligned}
\end{equation}
Here, $\delta$ denotes the depth along the ray from the texel to the splat's plane. 
The local 2D basis $\bm{l}_k \in \mathbb{R}^{3 \times 2}$ is defined following ~\cite{huang20242d} as: $\bm{l}_k=\begin{pmatrix}
    [R(\bm{q}_k)]_{[1]} /{[\bm{s}_k]_{[1]}}, [R(\bm{q}_k)]_{[2]}/{[\bm{s}_k]_{[2]}}
\end{pmatrix}$.
Finally, the splat’s contribution (alpha value) at that texel is computed with the 2D Gaussian kernel:
\begin{equation}
    \alpha_{(j),k} = \exp\left(-\frac{1}{2}\left(\hat{\bm{x}}_{(j),k,[1]}^2 + \hat{{\bm{x}}}_{(j),k,[2]}^2\right)\right)\cdot o_k.
\end{equation}
After the alpha value of the splat for the texel $\bm{t}_{(j)}$ is determined, we obtain the values of displacement, surface normal, and color of the texel through volumetric rendering. 

    \begin{figure*}[t]
        \includegraphics[width=\linewidth]{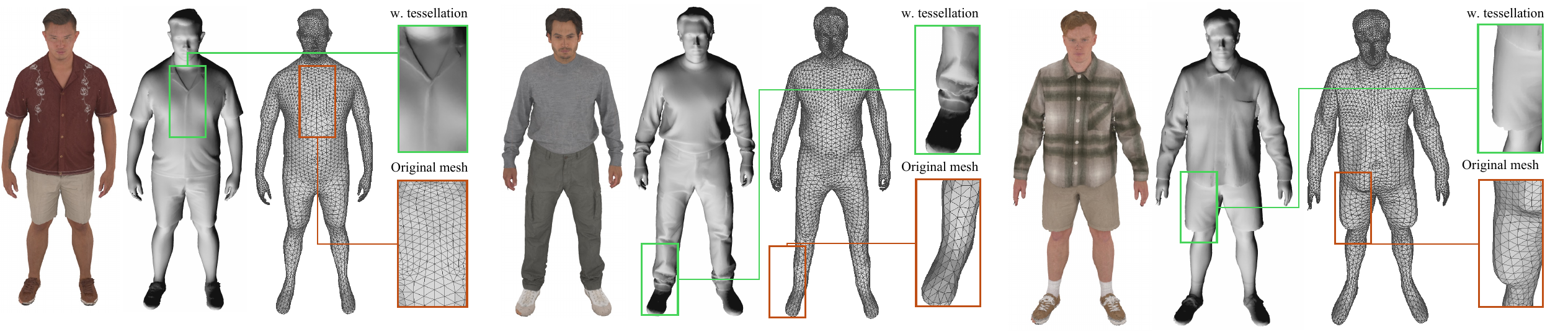}
        \captionof{figure}{
        OpenGL-rendered mesh results for avatar reconstruction (from left to right): rendering with diffuse map, rendering with displacement and normal maps, and the deformed mesh.
        Despite the low-poly geometry, normal and displacement maps recover fine surface details such as folds and cuffs.
        Avatar models \textcopyright{} Synthesia.
        }
        \Description{Avatar result}
        \label{fig:avatar}
    \end{figure*}

\subsection{Texture Map Refinement:}

Recent studies have pointed out the limitations of 3DGS in handling high-frequency textures~\cite{yu2024mip, xu2024supergaussians}. 
Our explicitly parameterized, resolution-bound texture map prevents bottlenecks and eliminates the need for costly pipeline tweaks to recover only marginal visual details.
Moreover, the commonly used spherical harmonics (SH) representation in GS encodes view-dependent color, which leads to slight inconsistencies when projecting splats onto a surface for texture extraction.


To further enhance texture quality, 
we add a lightweight texture refinement step after optimization. 
For each texel: 
\begin{enumerate}[itemsep=0pt, topsep=2pt]
  \item We map each texel’s UV coordinate to its 3D surface point $\hat{\bm{p}}$ using barycentric interpolation and then apply displacement $d$ along the corresponding face normal $\bm{n}$ to obtain its deformed position: $\hat{\bm p}^* = \hat{\bm{p}} + d\bm{n}$ (indices omitted for brevity). 
  \item We reproject \(\hat{\bm p}^*\) into image space with camera parameters $\pi$, thereby yielding $\bm{x}^\ast = \pi\bigl(\hat{\bm{p}}^\ast\bigr)$.
  \item We compare its reprojected depth $D^\ast$ to the splatting-rendered depth $D(\bm{x}^\ast)$. The point is considered visible if $ \bigl|D^* - D(\bm{x}^*)\bigr| < D_{\mathrm{th}}$ (we used $D_{\mathrm{th}} = 0.01$ in the experiments).
  \item For visible texels, we optimize their color by minimizing the $\ell_2$ loss between the texel value and the ground-truth image color at the projected position $\bm{x}^\ast$.
\end{enumerate}
This optimization requires only a few iterations to ensure a consistent texel color across all views.
Although we directly optimize texel colors here, this step is compatible with \rev{any differentiable reflectance formulation}, and can be used to recover intrinsic properties like albedo.

\subsection{Acceleration of Map Rasterization}

Rendering attribute maps per triangle is costly, given the size of the mesh.
To speed up this process, we assign one thread block per face and let the blocks rasterize all texels on their triangle in parallel.

For face $\bm{f}_j$, its block first loads all splats that are anchored to $\bm{f}_j$ and any faces up to three hops away; this ensures that the block includes splats that could affect neighboring triangles. 
In shared memory, the block stores the attributes of these splats and intermediate data, and then it sorts them according to descending depth $(\bm{p}_k - \hat{\bm{p}}_{(j)})\cdot \bm{n}_j$ so that blending proceeds from front to back. 
Finally, each thread intersects its texel with the sorted splats, applies the Gaussian projection model, and composites the results via volumetric alpha blending to produce the face’s attribute maps.



\section{Experiment}
We conducted two sets of experiments to evaluate DeMapGS.
\begin{table}[t]
    \centering
    \caption{Ablation study.  
    Variants: w/o 3DGS, 
    gradient diffusion (Diff),
    vertex re-alignment (Align),
    displacement (Disp), and tessellation (Tess).
}
    \small
    \begin{tabular}{|c|c|c|c|c|c|c|}
    \hline
         & Tree & Bust & Buddha & Book & Box & Cat \\\hline
       w/o 3DGS & 0.25 &1.43 &0.34 & 0.59 &  0.29& 1.12\\
       \rev{w/o 3DGS$^\ast$} & \rev{0.56} &\rev{1.88} &\rev{0.58} & \rev{1.09} &  \rev{0.76}& \rev{1.54}\\
       w/o Diff & 0.41 & 4.79& 0.87 & 10.46 &0.54 & 3.21\\
       w/o Align  & 0.24 &1.89 &0.29 & 0.51 & \textbf{0.28} & 1.39 \\
       w/o Disp  & 0.33 & 1.05 & 0.60 & 1.15 & 0.35 & 1.44\\
       w/o Tess & 0.25 & 0.83 & 0.28 & 0.49 & 0.30 & \textbf{1.00} \\
       full & \textbf{0.22} & \textbf{0.77} & \textbf{0.25} & \textbf{0.40} & 0.29 & \textbf{1.00}\\\hline
    \end{tabular}
    \label{tab:ablation}
\end{table}
    \noindent \textbf{Sketchfab 3D Scans:} 
    We used 3D scans from Sketchfab (CC-BY; bust: CC-BY-NC), and rendered 100 images per scene in Blender.
    We then used 80 images for training and 20 for testing.
    We used a subdivided cube as the initial template mesh.

    \noindent \textbf{ActorsHQ Dataset~\cite{icsik2023humanrf}:}
    We used a single static pose comprising 160 views for the training. 
    The SMPL-X model~\cite{pavlakos2019expressive}, which was initialized with the estimated pose, was used as the initial mesh template.
    To further enhance the surface detail, we utilized the reference normal maps obtained using Sapiens~\cite{khirodkar2025sapiens} as supervision targets for the normal consistency loss $\mathcal{L}_{\text{normal}}$.
    
\begin{figure}[t]
    \centering
    \includegraphics[width=\linewidth]{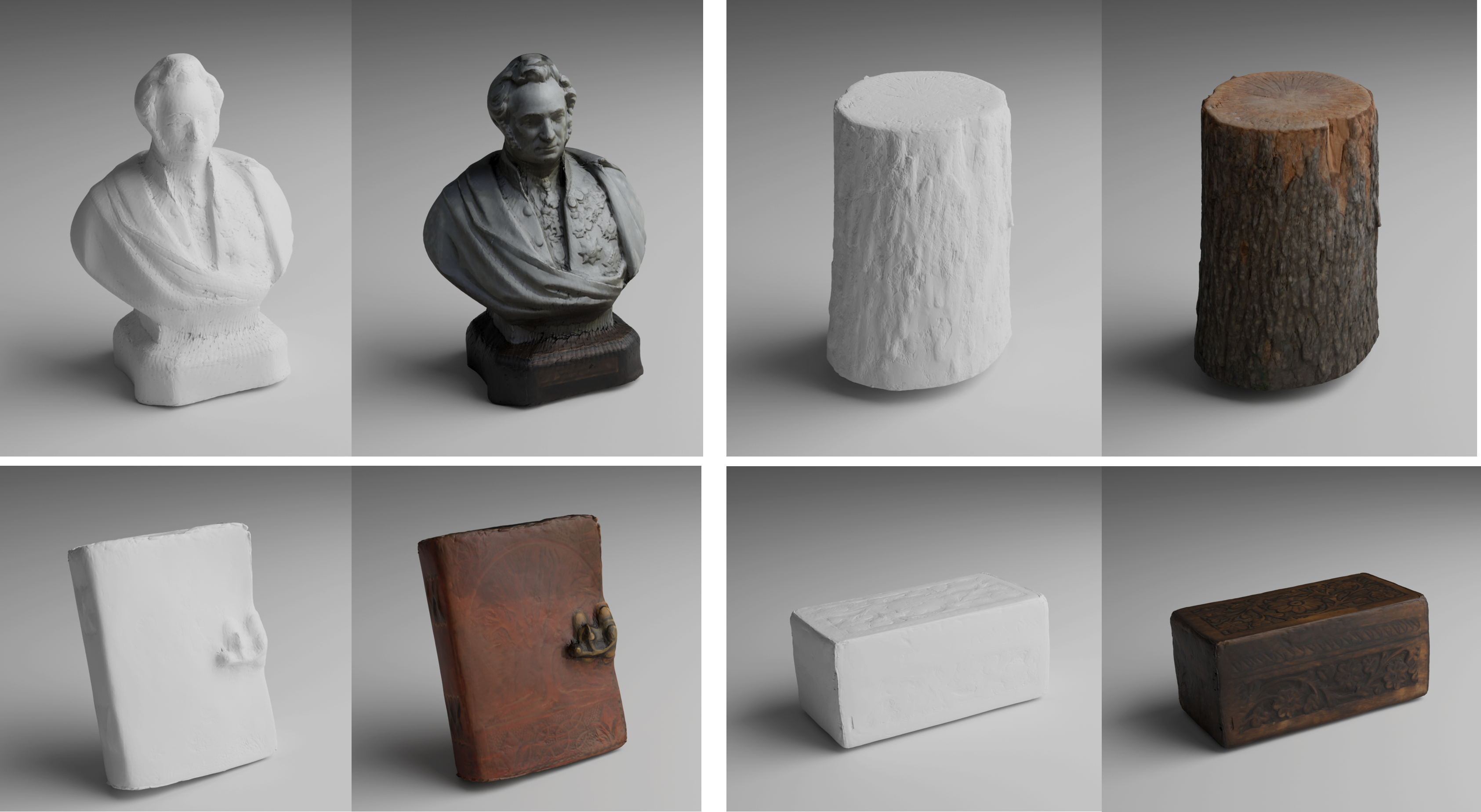}
    \caption{Rendered result in Blender using the deformed mesh and surface maps.}
    \label{fig:blender}
    \Description{Blender rendering}
\end{figure}
\subsection{Mesh Reconstruction Quality}
In our experiments, we present quantitative results on the Sketchfab 3D scan dataset and qualitative results on the ActorHQ dataset.
Since our method outputs a displacement map, it can be used for tessellation-based rendering.
However, standard geometric metrics cannot directly incorporate on-the-fly tessellation.
To approximate its effect during evaluation, we uniformly subdivided the mesh twice, resulting in a 16× increase in triangle count—equivalent to tessellation level 4.
We refer to this pre-subdivided version as \textbf{Ours-s} (static), while \textbf{Ours-d} uses actual dynamic tessellation at runtime (level 4 via OpenGL).

For comparative analysis, we focus on methods that prioritize mesh reconstruction quality over GS rendering fidelity. 
The comparison methods are categorized into two groups: reconstruction-based and deformation-based approaches.
The reconstruction-based methods include \textbf{SuGaR}~\cite{guedon2024sugar}, \textbf{2DGS}~\cite{huang20242d}, \textbf{GOF}~\cite{yu2024gaussian}, and \textbf{DG-Mesh}~\cite{liu2025dynamic}.
The deformable-based methods include \textbf{NRICP}~\cite{amberg2007optimal}, \textbf{Point2Mesh}~\cite{hanocka2020point2mesh} and \textbf{GaMeS}~\cite{waczynska2024games}.
We used 2DGS reconstructed mesh as supervision for \textbf{NRICP} and \textbf{Point2Mesh}.
    
In Table~\ref{tab:geom}, we use Chamfer Distance (CD1, CD2) to evaluate local geometric fidelity, and the L1 Sinkhorn distance (SD)~\cite{feydy2019interpolating} to assess the global distribution alignment.
Visualizations of the reconstructed meshes are shown in Fig.\ref{fig:geometry}.
Deformation-based methods generally exhibit unstable performance across different scenes, while reconstruction-based methods—especially 2DGS—tend to produce more consistent results. 
Despite being a deformation-based approach, our method significantly outperforms the other deformation-based baselines and achieves comparable geometric accuracy to 2DGS in most scenes.
One exception is the “cat” scene, where our method slightly underperforms, likely due to the highly concave regions that are difficult to reach via surface deformation.
\begin{figure}[t]
    \centering
    \includegraphics[width=0.9\linewidth]{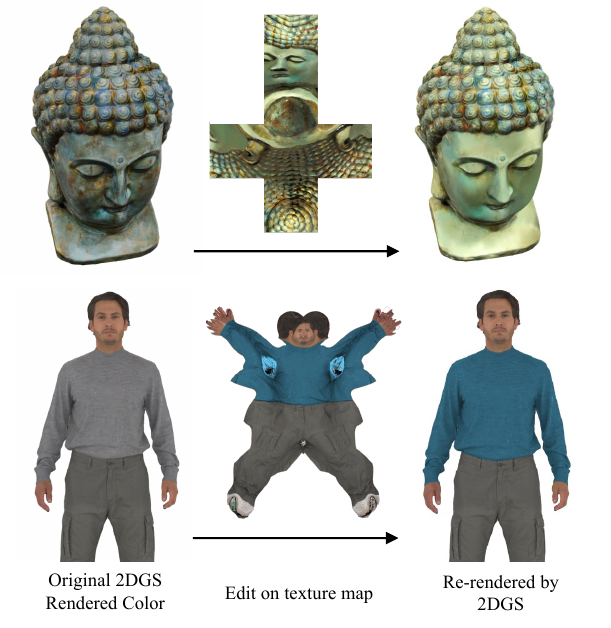}
    \caption{We first identify the modified UV pixels and compute their barycentric coordinates on the mesh. Using these coordinates, we locate the corresponding Gaussian splats that share similar barycentric anchors.
    The edited splats are then re-rendered using 2DGS.
}
\Description{Texture editing}
    \label{fig:texture-edit}
\end{figure}
\begin{figure}[t]
    \centering
    \includegraphics[width=\linewidth]{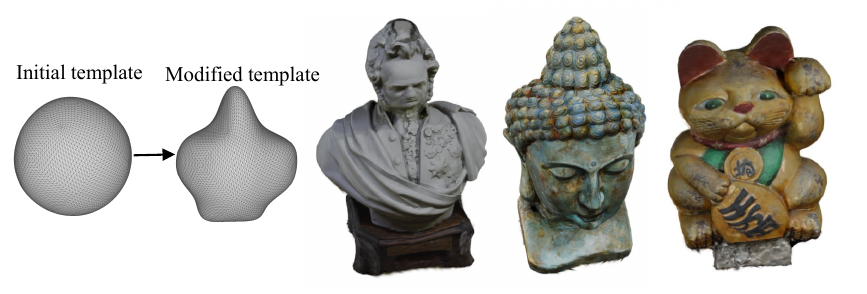}
    \vspace{-0.4cm}
    \caption{Editing Gaussian splats positions via template mesh deformation.
    We deform the initial mesh and compute vertex displacements, which are mapped to the target object by applying the global scale and rotation matrix $M$.
    The resulting vertex shifts are used to update the positions of the attached splats, which are then rendered using 2DGS.
    }
    \Description{Edit Geometry}
    \label{fig:deform-geom}
\end{figure}
\subsection{Mesh Rendering Performance}
Fig.\ref{fig:rgb-res} shows qualitative comparisons of rendered mesh colors for Sketchfab 3D scan data, and Table~\ref{tab:mesh} reports the rendering quality and efficiency of reconstruction-based methods.
PSNR and LPIPS~\cite{zhang2018perceptual} are computed from colored renderings of the reconstructed mesh, while FPS measures the rendering speed using OpenGL rasterization at a resolution of 1080$\times$1080. 

We observe that 2DGS produces noticeably lower-quality renderings due to its use of per-vertex colors. 
SuGaR achieves high visual quality using extremely dense meshes and UV-extracted textures, but often produces geometries with noisy or irregular surface artifacts.
Our method produces comparable results while being significantly more efficient: Ours-s renders at ~4× the FPS of SuGaR.
Ours-s and Ours-d achieve similar visual quality; Ours-d is slightly slower due to real-time tessellation, but still faster than SuGaR.

Fig.~\ref{fig:avatar} shows qualitative results on the ActorHQ dataset.
Since the initial template mesh is already well-aligned with the target geometry, our method produces high-quality reconstructions with detailed surface features.
As shown in the close-up regions, sharp creases and fine wrinkles—especially around joints or high-curvature areas—are difficult to capture on the SMPL-X template mesh alone.
The addition of normal maps and displacement-driven tessellation is crucial for recovering these high-frequency details.

\begin{figure}[t]
    \centering
    \includegraphics[width=\linewidth]{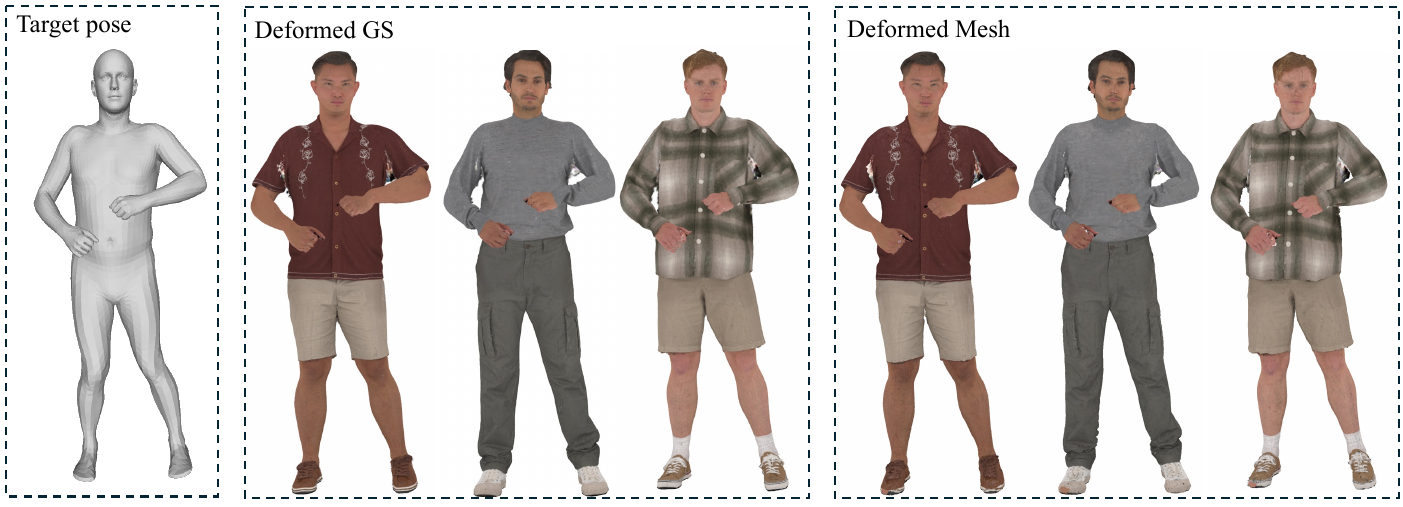}
    \caption{Pose-driven deformation of mesh and splats. We deform the avatar in Fig.~\ref{fig:avatar} to a target pose and update Gaussians using the same method as in Fig.~\ref{fig:deform-geom}.
    The rendered results show that the re-rendered mesh and re-rendered GS outputs are nearly identical. 
    Slight texture noise under the armpit appears due to missing texture coverage in the original data.}
    \label{fig:avatar-deform}
    \Description{Avatar animation}
\end{figure}
\begin{figure}[t]
    \centering
    \includegraphics[width=\linewidth]{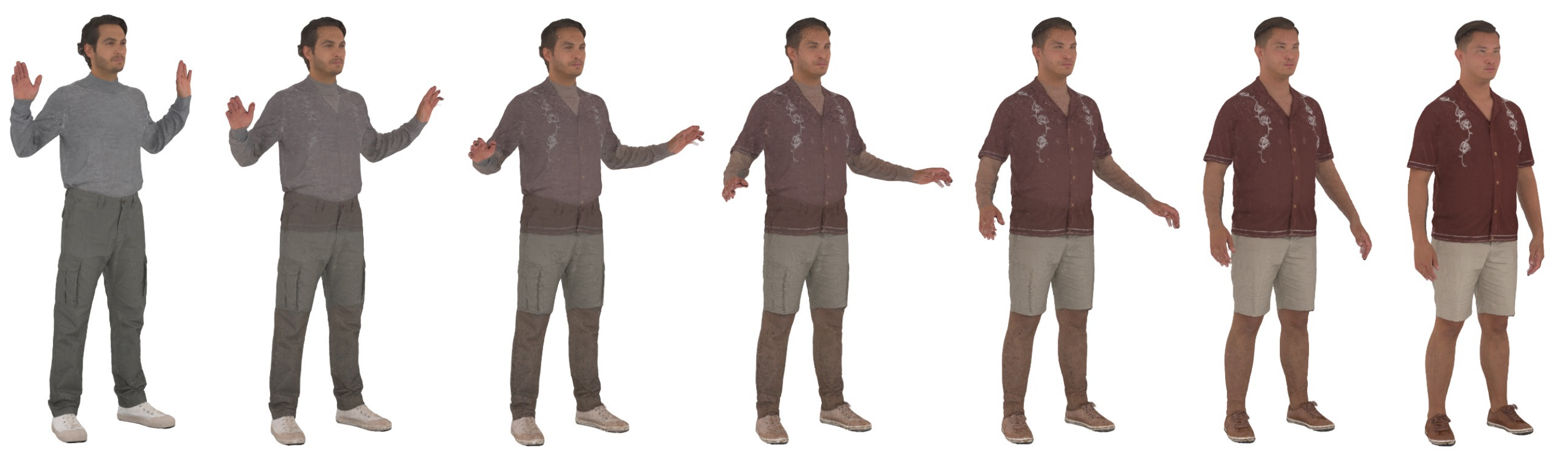}
    \vspace{-0.3cm}
    \caption
    {Interpolation between result meshes.
    We linearly interpolate all surface attributes between the two avatars.
    }
    \Description{Interpolating avarar}
    \label{fig:interpolate}
\end{figure}
\subsection{Ablation Studies}
We conducted ablation studies to assess the contribution of each proposed component.
Table~\ref{tab:ablation} reports the sum of CD1 and CD2 values for each variant.
\rev{Removing 3DGS degrades performance—where w/o 3DGS replaces the deformation-stage rasterizer to 2DGS with no extra 2DGS losses, while w/o 3DGS$^\ast$ also removes the final-stage 2DGS regularization, causing a larger drop.}
Disabling gradient diffusion (w/o Diff) leads to the largest drop in performance, indicating its key role in propagating supervision across the surface.
Removing vertex re-alignment (w/o Align) causes local geometry misalignment, especially near high-curvature regions.
We also observe performance degradation when omitting displacement maps (w/o Disp) or tessellation (w/o Tess), which together contribute to capturing fine surface detail.

\subsection{Applications}

Our deformed mesh with surface attribute maps enables editing not only at the vertex level, but also through surface-level modifications.
The resulting mesh can also be directly used in standard 3D software (e.g., Fig.~\ref{fig:blender}), making it practical for downstream applications.
Moreover, because our Gaussian splats are attached directly to the mesh surface, they inherit the same structure—allowing Gaussian-based representations to support editing and cross-object operations just like deformable surface models.

\noindent \textbf{Surface-based Editing: }
Explicit editing is straightforward for mesh surfaces with defined attribute maps. 
In our framework, such edits can be directly applied to the Gaussian splats as well. 
For example, Fig.~\ref{fig:texture-edit} shows a texture editing case.
Since our attribute extraction is differentiable, the edited splats can be further supervised to better match the desired appearance.
We also show example of geometry editing in Fig.~\ref{fig:deform-geom} and Fig.~\ref{fig:avatar-deform}.

\noindent \textbf{Cross-object Manipulation: }
Because all objects in our framework are reconstructed from a shared template, they reside in a unified canonical domain.  
This structure enables synchronized manipulation across multiple objects as in Fig.~\ref{fig:deform-geom} and Fig.~\ref{fig:avatar-deform}.
Moreover, our representation supports interpolation between different objects by linearly blending all the surface properties.  
This allows smooth transitions in both geometry and appearance, as demonstrated in the avatar interpolation example in Fig.~\ref{fig:interpolate}.

\section{Conclusion}
We presented a structured Gaussian splatting framework that anchors splats to a deformable mesh surface via surface-based parameterization, enabling joint optimization of mesh geometry and splat attributes.  
To support robust deformation, we introduced an optimization pipeline combining 2D/3D Gaussian Splatting rendering with gradient diffusion for large-step, regularized updates.  
We also proposed a postprocessing method to extract diffuse, displacement, and normal maps as standard UV textures.
Experiments demonstrate that our method achieves high-quality mesh reconstruction while preserving the photorealistic rendering of Gaussian splatting.  
Our structured representation further supports direct editing and cross-object manipulation of splats through shared surface correspondence.
A current limitation is the lack of topological flexibility, as our framework relies on a fixed mesh template.  
In future work, we plan to explore topology-adaptive extensions to support more complex deformations and geometry variations.
\nocite{ye2025gsplat}
\bibliographystyle{ACM-Reference-Format}
\bibliography{myref}

\clearpage
\appendix

\section{Camera-Space Derivatives and Non-degeneracy}
\label{app:derivatives-and-nondeg}
In this section, we prove that the camera-space positional gradient of a 3D Gaussian splat (3DGS) is full-directional—equivalently, it is not confined to any fixed plane. 
We first introduce notation and preliminaries (Sec.~\ref{sec:preliminary}); 
then (i) derive closed-form expressions for direction of position gradient (Secs.~\ref{sec:differential}–\ref{sec:xyz}), 
and (ii) establish non-degeneracy for splats with nonzero thickness (Sec.~\ref{sec:nondeg-r}). 
Finally, we remark that, under our formulation, the 2D splat or 3D splat without thickness limit degenerates to a fixed plane (Sec.~\ref{sec:2d}).

\subsection{Setup and Preliminary.}
\label{sec:preliminary}
A Gaussian splat center in camera space is $\bm{p}=(X,Y,Z)^\top$ with $Z>0$. Normalized image coordinates and depth
\[
u=\frac{X}{Z},\qquad v=\frac{Y}{Z},\qquad d=Z.
\]
Let the target pixel to render be $\bm{x}=(x,y)^\top$. Define residual on the normalized plane
\[
\bm{r}=\boldsymbol\mu-\bm{x},\qquad \boldsymbol\mu=(u,v)^\top\in\mathbb R^2.
\]
The (pinhole) projection Jacobian~\cite{ye2025gsplat} is
\[
J
=\frac{1}{d}
\begin{bmatrix}
1&0&-u\\[2pt]
0&1&-v
\end{bmatrix}\in\mathbb R^{2\times3}.
\]
Let the 3D covariance of the splat in the camera space be $\Sigma\in\mathbb R^{3\times3}$, which should be a rank $3$ and symmetric positive-definite. 
Its induced 2D image-plane covariance and inverse are
\[
\Lambda=J\Sigma J^\top\in\mathbb R^{2\times2}.
\]
The Gaussian function is defined by
\[
\mathcal{G}(\bm p, \bm x) = e^{-\frac12\,\bm{r}^\top \Lambda^{-1}\bm{r}}.
\]
For simplicity, we denote the exponential part of Gaussian function as
\[
E(\boldsymbol\mu,d)
=\frac12\,\bm{r}^\top \Lambda^{-1}\bm{r}.
\]
Since $\partial\mathcal L/\partial E$ is a scalar, the 3 dimensional direction of the positional gradient equals that of $\nabla_{\bm{p}}E$.

\subsection{Matrix-calculus identities and differentials}

\label{sec:differential}
\noindent In this section, we first derive the differential identity $dE$ for the Gaussian energy \(E=\tfrac12\,\bm r^\top\Lambda^{-1}\bm r\), which will be used in later calculations.

Two standard identities will be used:
\begin{equation}
\mathrm d(\Lambda^{-1})=-\Lambda^{-1}(\mathrm d\Lambda)\Lambda^{-1},
\qquad
\mathrm d\Lambda=(\mathrm dJ)\Sigma J^\top+J\Sigma(\mathrm dJ)^\top.
\label{eq:inv-and-dLambda}
\end{equation}
Differentiating $E=\frac12\,\bm{r}^\top\Lambda^{-1}\bm{r}$ gives
\begin{equation}
   \mathrm dE
=\tfrac12\big(\mathrm d\bm{r}^\top\,\Lambda^{-1}\bm{r}
+\bm{r}^\top\,\mathrm d(\Lambda^{-1})\,\bm{r}
+\bm{r}^\top\,\Lambda^{-1}\,\mathrm d\bm{r}\big).
\end{equation}
Define $\bm{g}:=\Lambda^{-1}\bm{r}\in\mathbb R^2$ and use the cyclic property of trace, we can write
\begin{align}
\mathrm dE
&=\bm{g}^\top\mathrm d\bm{r}-\tfrac12\,\bm{g}^\top(\mathrm d\Lambda)\,\bm{g} \nonumber\\
&=\bm{g}^\top\mathrm d\bm{r}
-\tfrac12\,\mathrm{tr}\!\big(\bm{g}\bm{g}^\top(\mathrm d\Lambda)\big).
\label{eq:dE-raw}
\end{align}
\begin{equation}
\mathrm{tr}\!\big(\bm{g}\bm{g}^\top(\mathrm d\Lambda)\big)=\mathrm{tr}\!\big(\bm{g}\bm{g}^\top(\mathrm dJ)\Sigma J^\top\big)
+\mathrm{tr}\!\big(\bm{g}\bm{g}^\top J\Sigma(\mathrm dJ)^\top\big)\nonumber.
\end{equation}
Since Frobenius inner product allows $\langle A,B\rangle_F=\mathrm{tr}(A^\top B)=\mathrm{tr}(A B^\top)$,
we have 
\begin{equation}
\begin{aligned}
\mathrm{tr}\!\big(\bm{g}\bm{g}^\top J\Sigma(\mathrm dJ)^\top\big)
&=\langle\bm{g}\bm{g}^\top J,\  \left(dJ\right)\Sigma\rangle_F\\
&=\mathrm{tr}\big(\left(\bm{g}\bm{g}^\top J\right)^\top (dJ)\Sigma\big)\\
&=\mathrm{tr}\big(J^\top \bm{g}\bm{g}^\top(dJ)\Sigma\big)=\mathrm{tr}\big( \bm{g}\bm{g}^\top(dJ)\Sigma J^\top\big)
\end{aligned}
\end{equation}
Defining $\bm{h}:=\Sigma J^\top\bm{g}$ and using the cyclic property, we can simplify it as
\begin{equation}
\mathrm{tr}\!\big(\bm{g}\bm{g}^\top J\Sigma(\mathrm dJ)^\top\big)=\,\langle \,\bm{g}\bm{h}^\top,\ \mathrm dJ\rangle_F=\bm{h}^\top(\mathrm{d} J)^\top\bm{g}.
\label{eq:trace-simplify}
\end{equation}
Substituting into \eqref{eq:dE-raw} yields the fundamental differential identity
\begin{equation}
\boxed{
\quad
\mathrm dE
=\bm{g}^\top\mathrm d\bm{r}
-\bm{h}^\top (\mathrm{d} J)^\top\bm{g}.
\quad}
\label{eq:dE-master}
\end{equation}

\subsection{First-order derivatives w.r.t.\ $(u,v,d)$}
\label{sec:uvd
}
In this section, using the differential identity $\mathrm dE=\bm{g}^\top\,\mathrm d\bm{r}-\bm{h}^\top(\mathrm dJ)^\top\bm{g}$ and the elementary partials of $\bm{r}$ and $J$ in \eqref{eq:partials-r-and-J}, we obtain the explicit first-order derivatives for $\partial_u \bm E, \partial_v \bm E$ and $\partial_d \bm E$.

Starting from the elementary partials of $\bm r$ and $J$:
\begin{equation}
\partial_u\bm{r}=\begin{bmatrix}
    1\\0
\end{bmatrix},\quad
\partial_v\bm{r}=\begin{bmatrix}
    0\\1
\end{bmatrix},\quad
\partial_d\bm{r}=\begin{bmatrix}
    0\\0
\end{bmatrix},
\end{equation}
\begin{equation}
\partial_u J=\tfrac{1}{d}\!\begin{bmatrix}0&0&-1\\[2pt]0&0&0\end{bmatrix},\quad
\partial_v J=\tfrac{1}{d}\!\begin{bmatrix}0&0&0\\[2pt]0&0&-1\end{bmatrix},\quad
\partial_d J=-\tfrac{1}{d}\,J.
\label{eq:partials-r-and-J}
\end{equation}
From \eqref{eq:dE-master} and denote $g_n$ as the $n$-th component of $\bm{g}$, $h_n$ as the $n^{\mathrm{th}}$ row value of $\bm{h}$, 
\begin{equation}
\boxed{
    \begin{aligned}
        \partial_u E
&=\bm{g}^\top(\partial_u\bm{r})-\bm{h}^\top(\partial_u J)^\top\bm{g}
= g_1-\frac{-h_3 g_1}{d}
= g_1\Big(1+\frac{h_3}{d}\Big),\\[4pt]
\partial_v E
&= g_2\Big(1+\frac{h_3}{d}\Big).\label{eq:dEv}
    \end{aligned}}
\end{equation}
For $d$,
\begin{equation}
\boxed{\partial_d E
= \frac{1}{d}\,\bm{h}^\top J^\top\bm{g}
=\frac{1}{d}\,\bm{r}^\top\bm{g}
=\frac{2}{d}\,E,}
\label{eq:dEd}
\end{equation}
where we used $\bm{h}^\top J^\top =\bm{g}^\top J \Sigma J^\top = \bm{g}^\top \Lambda$ and $\Lambda \bm{g} =\bm{r}$.

\subsection{Chain rule to camera coordinates $(X,Y,Z)$}
\label{sec:xyz}
Finally, we can get the partial derivatives of $\partial_XE, \partial_YE$ and $\partial_ZE$ by using chain rules from $(u, v, d)$ to $(X, Y, Z)$.

Since $(u,v,d)=(X/Z,\ Y/Z,\ Z)$, the Jacobian is
\[
T:=\frac{\partial(u,v,d)}{\partial(X,Y,Z)}
=\begin{bmatrix}
Z^{-1}&0&-u Z^{-1}\\[2pt]
0&Z^{-1}&-v Z^{-1}\\[2pt]
0&0&1
\end{bmatrix},\qquad \det T=\frac{1}{Z^2}>0.
\]
By the chain rule,
\begin{equation}
\nabla_{\bm{p}}E=
\begin{bmatrix}\partial_XE\\ \partial_YE\\ \partial_ZE\end{bmatrix}
=T^\top
\begin{bmatrix}\partial_uE\\ \partial_vE\\ \partial_dE\end{bmatrix}
=
\begin{bmatrix}
Z^{-1}\partial_uE\\[2pt]
Z^{-1}\partial_vE\\[2pt]
-\frac{u}{Z}\partial_uE-\frac{v}{Z}\partial_vE+\partial_dE
\end{bmatrix}.
\label{eq:gradXYZ}
\end{equation}
Combining \eqref{eq:dEv}--\eqref{eq:dEd} and using $d=Z$ yields the \emph{explicit} camera-space derivatives:
\begin{equation}
\boxed{\begin{aligned}
       \partial_XE&=\frac{g_1}{Z}\Big(1+\frac{h_3}{Z}\Big),\qquad
\partial_YE=\frac{g_2}{Z}\Big(1+\frac{h_3}{Z}\Big),\\
\partial_ZE&=\frac{1}{Z}\Big(2E-\big(u g_1+v g_2\big)\Big(1+\frac{h_3}{Z}\Big)\Big),
    \end{aligned}}
    \label{eq:XYZ-final} 
\end{equation}

\subsection{Non-degeneracy: position gradients does not lie in any fixed plane}
\label{sec:nondeg-r}
We prove non-degeneracy of $\nabla_{\bm p}E$ by contradiction: assume a fixed $\bm n=(\alpha,\beta,\gamma)^\top$ is orthogonal to $\nabla_{\bm p}E$ for all configurations. 
We reparameterize with the residual $\bm r=\boldsymbol{\mu}-\bm x$ (Sec.~\ref{sec:r}), express all terms in $\bm r$, and control non-explicit dependencies via $O(\|\bm r\|)$ expansions (Sec.~\ref{sec:or}). 
Substituting into the plane condition $\bm n^\top\nabla_{\bm p}E\equiv0$ (Sec.~\ref{sec:nondeg-r}), the inhomogeneous orders in $\bm r$ force first $\gamma=0$ (Sec.~\ref{sec:gamma}) and then $\alpha=\beta=0$ (Sec.~\ref{sec:alphabeta}). Hence no nonzero fixed $\bm n$ exists and $\nabla_{\bm p}E$ is not confined to any fixed plane.

\subsubsection{Algebraic identities in $r$.}
\label{sec:r}
Let
\[
\begin{aligned}
   & M(\boldsymbol \mu):=\begin{bmatrix}1&0&-u\\[2pt]0&1&-v\end{bmatrix},\qquad
J=\frac{1}{Z}M(\boldsymbol \mu),\\
&\Lambda(\boldsymbol\mu,Z)=J\Sigma J^\top=\frac{1}{Z^2}\,M\Sigma M^\top.
\end{aligned}
\]
Define the $2\times2$ symmetric positive definite matrix
\[
G(\boldsymbol\mu):=(M(\boldsymbol\mu)\Sigma M(\boldsymbol\mu)^\top)^{-1}\succ0,
\]
so that
\[
\Lambda^{-1}(\boldsymbol\mu,Z)=Z^2\,G(\boldsymbol\mu).
\]
Thus, with $\bm r=\boldsymbol\mu-\bm x$,
\begin{equation}
\bm g:=\Lambda^{-1}\bm r=Z^2\,G(\boldsymbol\mu)\,\bm r,\qquad
E=\tfrac12\,\bm r^\top\Lambda^{-1}\bm r=\tfrac{Z^2}{2}\,\bm r^\top G\boldsymbol(\mu)\,\bm r.
\label{eq:gE-in-r}
\end{equation}
For $h_3$, write $\bm h:=\Sigma J^\top\bm g$ and $e_3=(0,0,1)^\top$. Then
\[
\eta:=\frac{h_3}{Z}= \frac{e_3^\top \Sigma J^\top \bm g}{Z}
= \frac{1}{Z^2}\, \underbrace{\big(e_3^\top \Sigma M(\boldsymbol\mu)^\top\big)}_{=:~c(\boldsymbol \mu)^\top}\, \bm g
= c(\boldsymbol\mu)^\top G(\boldsymbol\mu)\,\bm r,
\]
where, with $\Sigma=(\sigma_{ij})$,
\begin{equation}
c(\boldsymbol\mu)
=\begin{bmatrix}\sigma_{31}-u\,\sigma_{33}\\[2pt]\sigma_{32}-v\,\sigma_{33}\end{bmatrix}
= c(\bm x)\;-\;\sigma_{33}\,\bm r,\qquad c(\bm x):=\begin{bmatrix}\sigma_{31}-x\sigma_{33}\\ \sigma_{32}-y\sigma_{33}\end{bmatrix}.
\label{eq:eta-in-r}
\end{equation}
Finally,
\begin{equation}
u\,g_1+v\,g_2
= (x+r_1)g_1+(y+r_2)g_2
= Z^2\big([x,y]\,G(\boldsymbol\mu)\,\bm r\;+\;\bm r^\top G(\boldsymbol\mu)\,\bm r\big),
\label{eq:ug1vg2-in-r}
\end{equation}
and $2E=Z^2\,\bm r^\top G(\boldsymbol\mu)\,\bm r$ by \eqref{eq:gE-in-r}.

\subsubsection{$\|G(\boldsymbol\mu)-G(\bm x)\|$ is $O(\|\bm r\|)$.}
\label{sec:or}
Fix $\bm x$ and write $\boldsymbol\mu=\bm x+\bm r$ with $\varepsilon:=\|\bm r\|$. Let $M(\boldsymbol\mu)=M_0+\Delta M$ where $M_0:=M(\bm x)$ and $\Delta M=\begin{bmatrix}0&0&-r_1\\0&0&-r_2\end{bmatrix}$, so $\|\Delta M\|=\varepsilon$. Set $K(\boldsymbol\mu):=M(\boldsymbol\mu)\Sigma M(\boldsymbol\mu)^\top$ and $K_0:=K(\bm x)$; then
\[
\Delta K:=K(\boldsymbol\mu)-K_0=\Delta M\,\Sigma M_0^\top+M_0\,\Sigma\,\Delta M^\top+\Delta M\,\Sigma\,\Delta M^\top,
\]
hence $\|\Delta K\|\le 2\|\Sigma\|\,\|M_0\|\,\varepsilon+\|\Sigma\|\,\varepsilon^2$. Let $m:=\lambda_{\min}(K_0)>0$ and take $\varepsilon$ small so that $\|\Delta K\|\le m/2$. With $G(\boldsymbol\mu)=K(\boldsymbol\mu)^{-1}$ and $G_0:=K_0^{-1}$,
\[
G(\boldsymbol\mu)-G_0
=K_0^{-1}(K_0-K(\boldsymbol\mu))K(\boldsymbol\mu)^{-1},
\]
so $\|G(\boldsymbol\mu)-G_0\|\le \|G_0\|\,\|\Delta K\|\,\|K(\boldsymbol\mu)^{-1}\|\le (1/m)\cdot\|\Delta K\|\cdot(2/m)=O(\varepsilon)$ by Weyl’s inequality. Therefore $G(\boldsymbol\mu)=G(\bm x)+O(\|\bm r\|)$ as $\|\bm r\|\to0$.

\subsubsection{Plane condition in $\bm r$.}
\label{sec:plane}
Let $\bm n=(\alpha,\beta,\gamma)^\top$ be constant and suppose
$\bm n^\top\nabla_{\bm p}E\equiv0$ on $\Omega$. Using
\[
\begin{aligned}
    \partial_XE&=\tfrac{1}{Z}\,g_1(1+\eta),\quad
\partial_YE=\tfrac{1}{Z}\,g_2(1+\eta),\\
\partial_ZE&=\tfrac{1}{Z}\Big(2E-(u g_1+v g_2)(1+\eta)\Big),
\end{aligned}
\]
and substituting \eqref{eq:gE-in-r}--\eqref{eq:ug1vg2-in-r}, multiplying by $Z>0$ and then by $Z^{-2}$, we obtain the \emph{$r$-form} of the plane condition:
\begin{equation}
\begin{aligned}
   (1+\eta)\,\Big(
   \begin{bmatrix}
  \alpha\\\beta     
   \end{bmatrix}^\top G(\boldsymbol\mu)\,\bm r-\gamma\big(\begin{bmatrix}
  x\\y   
\end{bmatrix}^\top G(\boldsymbol\mu)\,\bm r+ & \bm r^\top G(\boldsymbol\mu)\,\bm r\big)\Big)
\;\\&+\;\gamma\,\bm r^\top G(\boldsymbol\mu)\,\bm r
=0. 
\label{eq:plane-r}
\end{aligned}
\end{equation}

\subsubsection{Step 1: $\gamma=0$.}
\label{sec:gamma}
Let $\bm r=\varepsilon \bm t$ with fixed $\|\bm t\|=1, \;\bm t\in\mathbb R^2$ and $\varepsilon\downarrow0$. Since $G(\boldsymbol\mu)=G(\bm x)+O(\varepsilon)=G_0+O(\varepsilon)$ and $\eta=c(\boldsymbol\mu)^\top G(\boldsymbol\mu)r=O(\varepsilon)$, the leading ($O(\varepsilon)$) terms of \eqref{eq:plane-r} give
\[
\big([\alpha,\beta]-\gamma[x,y]\big)\,G_0\,\bm t\;=\;0\quad\forall\, \bm t
\ \Longrightarrow\ 
[\alpha,\beta]=\gamma[x,y].
\]
Plugging this back into \eqref{eq:plane-r} cancels the linear terms and yields the exact reduction
\[
-\gamma\,\eta\,\bm r^\top G(\boldsymbol\mu)\,\bm r=\mathbf 0.
\]
Since $G(\boldsymbol\mu)\succ0$, we have $\bm r^\top G(\boldsymbol\mu)\,\bm r>0$ for $r\neq\mathbf 0$. Moreover, by \eqref{eq:eta-in-r}, $\eta$ is not identically zero on any neighborhood of $\bm r=\mathbf 0$ (if $c(\bm x)\neq\mathbf 0$, then $\eta$ has a linear term $c(\bm x)^\top G_0\,\bm r$; if $c(\bm x)=\mathbf 0$, then $\eta=-\sigma_{33}\,\bm r^\top G_0 \bm r+O(\|\bm r\|^3)\not\equiv0$ since $\sigma_{33}>0$). Hence the identity above forces
\[
\boxed{\ \gamma=0.\ }
\]

\subsubsection{Step 2: $\alpha=\beta=0$.}
\label{sec:alphabeta}
With $\gamma=0$, \eqref{eq:plane-r} reduces to
\[
(1+\eta)\,[\alpha,\beta]\,G(\boldsymbol\mu)\,\bm r=0
\]
Taking $\varepsilon$ small ensures $1+\eta\neq0$, hence $[\alpha,\beta]\,G(\boldsymbol\mu)\,\bm r\equiv0$. Passing to the limit $\boldsymbol\mu\to\bm x$ gives $[\alpha,\beta]\,G_0\,\bm r\equiv0$ for all $\bm r$, so (invertibility of $G_0$)
\[
\boxed{\ \alpha=\beta=0.\ }
\]

\subsubsection{Conclusion.}
\label{sec:conclu}
We have shown that the only constant $\bm n=(\alpha,\beta,\gamma)$ satisfying $\bm n^\top\nabla_{\bm p}E\equiv0$ on $\Omega$ is $\bm n=\mathbf0$. Therefore, $\nabla_{\bm p}E$ does not lie in any fixed plane.

\subsection{Remark: 2DGS degeneracy}
\label{sec:2d}
In the 2DGS setting the splat has no depth variance or cross-covariance, i.e.
\[
\Sigma_{13}=\Sigma_{23}=\Sigma_{33}=0\quad.
\]
Hence
\[
c(\boldsymbol\mu)=\begin{bmatrix}\Sigma_{31}-u\Sigma_{33}\\ \Sigma_{32}-v\Sigma_{33}\end{bmatrix}=\mathbf{0}
\quad\Rightarrow\quad
h_3=0.
\]
Then plane condition reduces to
\[
\alpha g_1+\beta g_2+\gamma\big(2E-ug_1-vg_2\big)=0
\ \Longleftrightarrow\
(\alpha-\gamma x)g_1+(\beta-\gamma y)g_2=0,
\]
since \(2E=(u-x)g_1+(v-y)g_2\).
Choosing any nonzero \(n\) proportional to \((x,y,1)^\top\) (e.g. \(\bm n=(x,y,1)^\top\)) makes the identity hold on \(\Omega\).
Thus there exists a constant \(\bm n\neq\mathbf{0}\) with \(\bm n^\top\nabla_{\bm{p}}E\equiv0\); the camera-space positional
gradient lies in the fixed plane \(\bm n^\perp\) and is effectively two-dimensional in the 2DGS case.

\end{document}